\documentclass[1p]{elsarticle}
\journal{journal}

\usepackage{cite}
\usepackage{hyperref}

\usepackage{amsmath}    % For mathematical symbols and equations
\usepackage{amssymb}    % For additional mathematical symbols
\usepackage{graphicx}   % For including images
\usepackage{geometry}   % For customizing page layout
\usepackage{setspace}   % For setting line spacing
\usepackage{booktabs}   % For improved table formatting
\usepackage{enumitem}   % For customizing lists
\usepackage{mathptmx} % Times font for both Math and texts

\usepackage{xcolor} % For colors
\usepackage{soul}% for highlight in yellow using \hl{This text is highlighted in yellow.}

\definecolor{myblue}{RGB}{0,0,255}
\hypersetup{
  colorlinks=true,
  linkcolor=blue,
  citecolor=blue,
  urlcolor=blue,
}

\begin{document}

\title{Low-energy neutrino responses for ${}^{71}$Ga by electron capture rates, \\
charge exchange reactions and shell model calculations}

\author[first]{Yoritaka Iwata \corref{cor1}}
\ead{y-iwata@s.keiho-u.ac.jp}

\author[second]{Hiroyasu Ejiri}

% \ead{\href{mailto:ejiri@rcnp.osaka-u.ac.jp}{ejiri@rcnp.osaka-u.ac.jp}}

\author[third]{Shahariar Sarkar}

\affiliation[first]{organization={Osaka University of Economics and Law},
            addressline={Yao},
            postcode={Osaka 581-0853},
            country={Japan}}

\affiliation[second]{organization={Research Center for Nuclear Physics, Osaka University},
            addressline={Osaka},
            postcode={567-0047},
            country={Japan}}

\affiliation[third]{organization={Department of Physics, Indian Institute of Technology Roorkee},
            addressline={Roorkee},
            postcode={247667},
            country={India}}

\cortext[cor1]{Corresponding author}

\begin{frontmatter}

\begin{abstract}
Weak Gamow-Teller (GT) responses for low-lying states in ${}^{71}\mathrm{Ga}$ are crucial for studying low-energy solar neutrinos and the Ga anomaly, i.e., the possible transition to the sterile state. The responses for the ground state, the first excited state, and the second excited state are evaluated for the first time using the experimental electron capture rates, the experimental charge exchange reaction (CER) rates corrected for the tensor-interaction effect and the theoretical interacting shell model (ISM) calculations. The contributions from the two excited states to the solar and ${}^{51}\mathrm{Cr}$ neutrinos are found to be $4.2 \pm 1.2\%$ of that for the ground state. This is slightly larger than the ISM values but little smaller than the CER values without corrections for the tensor interaction effect. The Ga anomaly is far beyond the uncertainty of the obtained nuclear responses.
\end{abstract}


\end{frontmatter}

\section{Introduction}
Neutrino ($\nu$) nuclear responses for the ${ }^{71} \mathrm{Ga}$ nucleus have recently been sharpened by the Ga anomaly \citep{Abdurashitov2006, Kaether2010, Giunti2011, Barinov2022a, Barinov2022b} and the possible transition to the sterile state \citep{Giunti2019, Basgupta2021, Giunti2021}. In fact, the ${ }^{71} \mathrm{Ga}$ isotopes have been used widely to study low-energy solar $\nu\mathrm{s}$ and low-energy $\beta$-decay $\nu\mathrm{s}$ from radio-active nuclei because of the low-energy thresh-hold energy of $E_{l h} \approx 232 \mathrm{keV}$ and the muti10 ton scale Ga detectors. Recently the BEST experiment using the two-sector (inner and outer sections) Ga detector has reported deficient of the detected $\nu s$ by $17-18 \%$ from the expected $\nu s$ \citep{Barinov2022a, Barinov2022b}, suggesting electron-neutrino transitions to the eV sterile state.

\begin{figure}[h]
    \centering
    \includegraphics[width=0.5\textwidth]{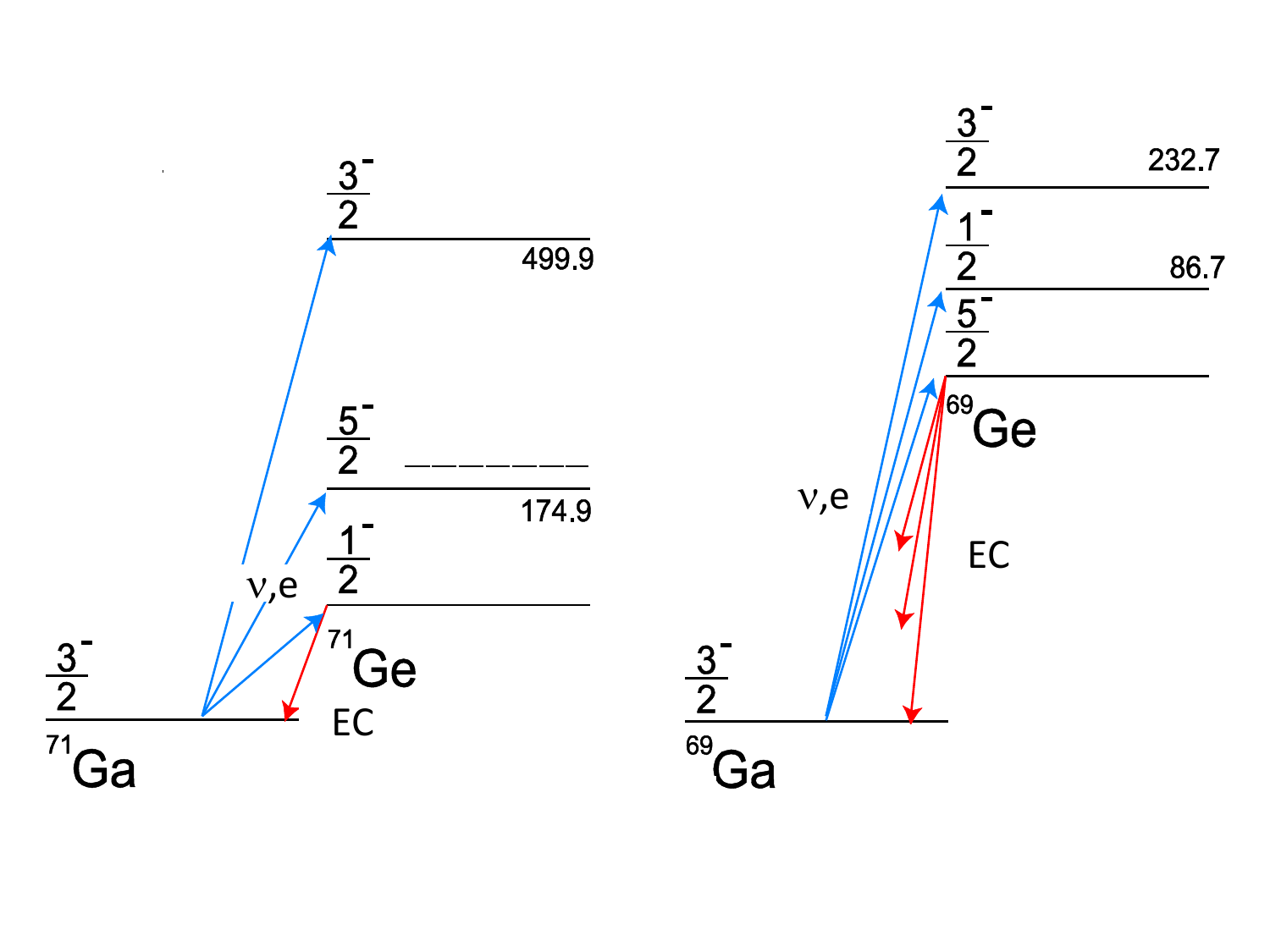}
    \caption{Level scheme for $\nu$ CC interactions and CERs on $^{71} \mathrm{Ga}$ and $^{69} \mathrm{Ga}$. Blue lines are for CCs. Red lines are for ECs.}
    \label{fig:level_scheme}
\end{figure}

One of key ingredients of the Ga experiment is the charged current (CC) $\nu$ response for the low-energy $\nu$s below 1 MeV. The nuclear states involved in the charged current $\nu$-interaction are the ground state with $1 / 2^{-}$, the first excited 175 keV state with $5 / 2^{-}$and the second excited 500 keV state with $J=3 / 2^{-}$. The level schemes for the CC interactions and the electron captures (EC) for $^{71} \mathrm{Ga}-^{71} \mathrm{Ge}$ are shown in Fig. \ref{fig:level_scheme}.

The $\nu$ interaction rate for the state $i~(i=0,1,2$ for the ground, the 1st and the second excited states) is 

\begin{equation}
R_{i}=f(\nu) G_{i}(\nu) g_{A}^{2} B_{i}(\mathrm{GT})
\end{equation}
where $f(\nu), G_{i}(\nu), g_{A}$ and $B_{i}(\mathrm{GT})$ are the $\nu$ flux, the phase space factor, the axial vector weak coupling for a free nucleon and the GT (Gamow-Teller) $\nu$-response, respectively. We use $g_{A}=1.27$ in units of the vector coupling of $g_{V}$. The response is expressed as $B_{i}(\mathrm{GT})=M_{i}(\mathrm{GT})^{2} /(2 J+1)$ with $M_{i}(\mathrm{GT})$ being the nuclear matrix element (NME) and $J=3 / 2$ being the ${ }^{71} \mathrm{Ga}$ ground state spin. The total interaction rate is given by the sum $\Sigma_{i} R_i$. The interaction operator is $\mathrm{T}(\mathrm{GT})=\tau \sigma$ with $\tau$ and $\sigma$ being the isospin and spin, respectively.

The present letter addresses the $\nu$-responses of $B_{i}(\mathrm{GT})$ for the low-lying 3 states. The responses relate the $\nu$-flux to the $\nu$-signal rate, and thus the accurate values for them are indispensable for studying the neutrino oscillation parameters among the three generations and the possible transition to the sterile-state, and also for studying the solar neutrinos.

The $B_0$(GT) for the ground-state ($i=0$) is given experimentally by the EC rate. It is a simple single quasi-particle (QP) transition of 2p(1/2)-2p(3/2). On the other hand, the transitions to the first ($i=1$) and second ($i=2$) excited states are not simple QP transition, but include the $l$-forbidden ($\Delta l=2$) transition of $2p(3/2)-1f(5/2)$ and the quadrupole-phonon transition. 
Then both theoretical and experimental studies for them are necessary for accurate evaluations of the $v$-responses. Nuclear responses (squares of nuclear matrix elements NMEs) for $v$-studies in nuclei have been discussed in review articles and references therein \citep{Ejiri2000,Ejiri2019}.

Charge exchange reactions (CERs) are powerful to study $v$ nuclear responses. The low energy-resolution (p,n) CER on $^{71}\text{Ga}$ was used to get the $B_2(GT)$ for the 2nd excited state \citep{Krofcheck1985, Bahcall1997}. Since the excitation to the 1st excited state was not resolved from the background, the half of the upper limit was assumed to be the $B_1(GT)$ \citep{Bahcall1997}. The high energy-resolution (${}^3\text{He}$, $t$) CERs on $^{69,71}\text{Ga}$ at RCNP (Research Center for Nuclear Physics) Osaka University were used to extract the responses $B_i$(GT) for the 3 states of $i=0, 1, 2$ \citep{Ejiri1998,Frekers2011}. Note that the CER proceeds by the central GT operator of $T$(GT) and the tensor one of $T(T)$ \citep{Austin1994,Fayache1997}, the latter of which gets significant in the present $i=1$ and $2$ cases \citep{Haxton1998}. 
Actually, CERs so far have been extensively used to get $B$(GT) by assuming only the GT transition operator, and thus limited to simple GT states with large  $B(\rm GT)$ . In the present work we show for the first time how to correct the CER data for the $T(\rm T)$ contribution.

Recently, an interacting shell model (ISM) calculation was made on both the GT and the tensor NMEs for $^{71}\text{Ga}$ \citep{Kostensalo2019}. The actual GT NMEs, however, are much reduced  compared with the shell model calculations because of the quenching effect associated with  non-nucleonic and other correlations beyond the ISMs and the transitions to the 1st and 2nd states are not simple OP ones, but include quadrupole phonon and $l$-forbidden ones, and thus accurate ISM evaluations for the GT NMEs are very hard. Thus experimental imputs certainly help for evaluating the realistic NMEs.

In the present work we evaluate for the first time the $\nu$-responses of $B_i$(GT) with $i=0,1,2$ for $^{71}$Ga by using fully the high energy resolution $(^{3}\text{He},t)$ CER data on $^{71,69}$Ga with appropriate correction for the tensor contributions of $T$(T), the EC data on $^{71,69}$Ga and also the interacting shell model calculations on the GT and tensor NMEs for $^{71,69}$Ga.

%=======================
\section{GT NMEs from electron captures, $\left(^{3} \mathrm{He}, t\right)$ CERs and shell-model calculations}
Experimental data to be used for evaluating the $\nu$ responses for the low lying states in $^{71} \mathrm{Ga}$ are the EC rates and the CER cross sections for $^{69,71}$Ga. The ISM calculations for them are made by using the \textbf{JUN45} interaction \citep{Honma2009}. The shell model calculations are performed in the $fpg$-shell model space, which includes the $p_{3/2}$, $f_{5/2}$, $p_{1/2}$, and $g_{9/2}$ orbitals. The JUN45 interaction is a realistic effective interaction derived from the Bonn-C potential, adjusted to fit experimental data in this mass region. This interaction has been shown to provide a good description of the spectroscopic properties of nuclei around $A=70$, enabling adequate computation of nuclear wave functions and transition matrix elements for beta-decay and neutrino processes.

\begin{figure}[h]
    \centering
    \includegraphics[width=0.6\textwidth]{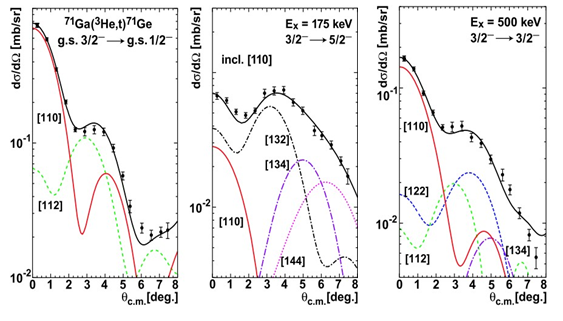}
    \caption{The CER angular distributions for the low-lying final states with spins of $J_{f}=1 / 2,3 / 2$ and $5 / 2$ in ${ }^{71} \mathrm{Ga}$ \citep{Frekers2011}. $\left[J_{p} J_{t} J_{r}\right]$ stands for the DWBA fits with the angular momentum transfers for the projectile (p), the target (t) and the relative momentum (r).}
    \label{fig:cer_angular}
\end{figure}

The EC NME is $M(\mathrm{ECGT})=0.585$ for the ground state transition of $^{71} \mathrm{Ga}$ $(3 / 2) \rightarrow{ }^{71} \mathrm{Ge}(1 / 2)$ \citep{Nakamura2010}. This is a typical single QP NME. The ISM NME is $M($ISMGT$)=0.791$. Thus the experimental NME is quenched with respect to the ISM NME by $k=g_{A}^{eff} / g_{A}=0.74$ with $g_{A}^{eff}$ being the effective axial-vector coupling. This is consistent with the quenching coefficient in medium heavy nuclei \citep{Ejiri2000, Ejiri2019}. The coefficient reflects such non-nucleonic (isobars and others) correlations and nuclear medium effects that are not explicitly included in the ISM calculation.

The NMEs of $k M$(ISMGT) for the low-lying 3 states are shown in Table \ref{tab:nme}. Then we use this quenching coefficient for other GT ISM NMEs in the $^{69,71} \mathrm{Ga}$ since the quenching effect is due to the nuclear core/medium and is common for low-lying states characteristic of the valence (surface) nucleon configurations. The experimental $M_{\beta}(\mathrm{GT})=0.088$ for the ground state transition of $^{69} \mathrm{Ga}(3/2) \rightarrow^{69} \mathrm{Ge}(5/2)$ \citep{Nesaraja2014} is almost an order of magnitude smaller than a typical single QP NME, indicating the transition is mainly a $l$-forbidden (p-wave $\rightarrow$ f-wave) transition.

\begin{table}[h]
    \centering
    \begin{tabular}{|c|c|c|c|c|c|}
        \hline
        $E_{f}(\mathrm{MeV})$ & $J_{f}$ & $M_{i}(110)$ & $k M_{i}(\mathrm{ISMGT})$ & $k M_{i}(\mathrm{ISMT1})$ & $M_{i}(\mathrm{GT})$ \\
        \hline
        0 & $1 / 2$ & $0.575 \pm 0.015$ & 0.585 & 0.104 & $0.603 \pm 0.02$\\
        0.175 & $5 / 2$ & $0.115 \pm 0.014$ & 0.107 & 0.454 & $0.010 \pm 0.05$\\
        0.500 & $3 / 2$ & $0.260 \pm 0.016$ & 0.071 & 0.017 & $0.255 \pm 0.03$ \\
        \hline
    \end{tabular}
    \caption{NMEs $M_{i}$ for the low-lying states of $i=0,1,2$ in ${ }^{71} \mathrm{Ge} .  E_{f}(\mathrm{MeV})$ : excitation energy, $J_{f}$ : spin. $M(110)$ : the absolute value of the CER NME. $k M$(ISMGT): ISM spin (GT) NME. $k M$(ISMT1): ISM T1 NME. $M(\mathrm{GT})$ : absolute value of GT NME. $k=0.74$ is the axial-vector quenching coefficient of $g_{\rm A}^{eff}/g_{A}$ for SM NMEs. Note $k M_{0}(\mathrm{SMGT})=M_{0}(\mathrm{ECGT})$. See text.}
    \label{tab:nme}
\end{table}

The ($^{3} \mathrm{He}, t$ ) CERs for $^{71,69} \mathrm{Ga}$ were studied at RCNP Osaka University by using the medium energy $^{3} \mathrm{He}$ projectile with $E=0.42 \mathrm{GeV}$, where the GT and T modes are preferentially excited because of the large axial-vector interaction at the energy \citep{Ejiri1998, Frekers2011}. The three states with $i=0,1,2$ are well resolved by using the high energy-resolution spectrometer. There the data were analyzed in the previous works \citep{Ejiri1998, Frekers2011} by using only the GT interaction operator of T(GT). 
as in most cases so far.   As discussed earlier, CER does include the tensor interaction, which should be well included in the non-quasi-particle  GT transitions. 
So we re-analyze them in the present work by including explicitly the tensor interaction T1 with rank 1 as well as the GT one.

In the shell model, the tensor operator $\mathrm{T1} = \tau[Y_2 \times \sigma]_1$ is evaluated using the same wave functions as the GT operator $\mathrm{T(GT)} = \tau \sigma$. The resulting tensor matrix element $M(\mathrm{ISMTI})$ quantifies the contribution of the tensor interaction to the CER cross section, modeled as $M_i(110) = M_i(\mathrm{GT}) + \varepsilon M_i(\mathrm{TI})$, where $\varepsilon$ is the mixing coefficient. By calculating $M(\mathrm{ISMTI})$ within the shell model, we correct the CER-derived $M_i(110)$ to extract the pure GT strength $M_i(\mathrm{GT})$, enhancing the precision of the neutrino response analysis.

The CER differential cross section for the final state $i$ is expressed as
\begin{equation}
\sigma_{i}=\sum_{\alpha} \sigma_{i}(\alpha), \quad \alpha=\left(J_{p}, J_{t}, J_{r}\right)
\end{equation}
where $\sigma_{i}(\alpha)$ is the $\alpha$ mode cross section to the state $i$. The transition mode is characterized by the transferred spins as $\alpha=J_{p}, J_{t}, J_{r}$ with $J_{p}, J_{t}, J_{r}$ being the projectile (p), target (t) and relative (r) spin transfers. The experimental cross sections for the low-lying three states and the DWBA (distorted wave Born approximation) fits are shown in Fig. \ref{fig:cer_angular}.

The cross section component relevant to the present GT weak response is $\sigma_{i}(110)$ with $J_{p}=1, J_{t}=1, J_{r}=0$. The CER triton shows the s-wave ( $J_{r}=0$ ) angular distribution of the square of the spherical Bessel function of $j_{0}(q r)$ with $q$ and $r$ being the momentum transfer and the effective interaction radius. The 110-mode cross section is expressed as
\begin{equation}
\sigma_{i}(110)=K\left[J_{\sigma\tau}\right]^{2} N B_{i}(110), 
\end{equation}
where $K$, $J_{\sigma\tau}$, $N$, and $B_{i}(110)$ are the kinematic factor, the interaction volume integral, the distortion factor, and the reduced 110 strength, respectively. Note that $K$ and $N$ are common for all three states since the energies of the out-going tritons are nearly same for the three states.

According to the standard way of extracting the reduced width of $B_{i}(110)$, we take the cross section ratio to the cross section $\sigma_{IA}(000)$ for the IAS (isobaric analogue state) with the known reduced width of $B_{IA}(000)=N-Z$ with $N, Z$ being the neutron and proton numbers. The IAS cross section is expressed as $\sigma_{IA}=K_{IA}\left[J_{\tau}\right]^{2} N_{IA} B_{IA}(000)$. Using $\left[J_{\sigma\tau}\right]^{2} /\left[J_{\tau}\right]^{2}=8.1$ \citep{Brown1981} and $B_{IA}(000)=9$, the reduced width $B_{i}(110)$ is expressed as $B_{i}(110)=1.03 \pm 0.05$ $\left(\sigma_{i}(110) / \sigma_{IA}(000)\right)$. The NME $M_{i}(110)$ is $(B_{i}(110) /(2 J+1))^{1 / 2}$ with $J=3 / 2$ being the initial state spin. The obtained NMEs of $M_{i}(110)$ are shown in Table \ref{tab:nme}. 
%\begin{figure}[h]
%    \centering
%    \includegraphics[width=0.9\textwidth]{figures/fig03.png}
  %  \caption{Absolute values of NMEs for low-lying three states with $J_{f}=1 / 2,3 / 2$ and $5 / 2$ in ${ }^{71}$ Ga. Left panel: GT NMEs $M_{i}(\mathrm{GT})$ (blue squares) derived from CER NMEs $M_{i}(110)$ by correcting for the tensor contributions and NMEs $M_{i}(110)$ (light blue triangles) from CER without correction for the tensor NME. Right panel: Shell model GT NMEs $k M_{i}(\mathrm{ISMGT})$ (blue squares) and shell model tensor NMEs $k M_{i}(\mathrm{ISMT1})$ (light blue triangles).}
  %  \label{fig:nme_plot}
%\end{figure}

The large $M_{0}(110)$ and the small $M_{1}(110)$ for the ground state and the 1st excited state suggest, respectively, the main GT component and the main non GT (tensor) component for the transitions to the ground state and the first excited state. The $M_{2}(110)$ for the second excited state is about a half of the $M_{0}(110)$, suggesting the small GT NME for this state.

The $M_{i}(110)$ is expressed as the sum of the GT and T1 NMEs \citep{Haxton1998},
\begin{equation}
M_{i}(110)=M_{i}(\mathrm{GT})+\epsilon M_{i}(\mathrm{T} 1)
\end{equation}
where $M_{i}(\mathrm{GT})$ is the GT NME for the GT operator $\tau \sigma$, and $M_{i}(\mathrm{~T} 1)$ is the T1 NME for the T1 operator $\tau\left[Y_{2}\times \sigma \right]_{1}$, and $\epsilon$ is the mixing coefficient. The higher rank $\mathrm{T} 2=\tau\left[Y_{2} \times \sigma\right]_{2}$ and $\mathrm{T} 3=\tau\left[Y_{2} \times\sigma \right]_{3}$ are involved in the cross sections of $\sigma(122)$ and $\sigma(132)$, where the outgoing particles are of d-wave tritons. Then they show up at the scattering angles of $\theta \approx 2-4 
 ~\mathrm{deg}$., but not at the forward angles, as shown in Fig. \ref{fig:cer_angular}.

The tensor NMEs $M$ (TL) are higher-order terms in cases of lepton-photon $\beta-\gamma$ decays and thus are ignored. They, however, get significant in case of the hadronic CERs, depending on the transition involved in the CERs. In case that the transition is mainly a simple spin-flip transition of $l \pm 1 / 2 \leftrightarrow l \mp 1 / 2, M(GT) \gg M(T1)$, while in case that the transition is mainly of $l$ forbidden transition of $l+1 / 2 \leftrightarrow(l+2)-1 / 2, M(\mathrm{~T} 1) \gg M(\mathrm{GT})$.

The observed CER cross sections show that $\sigma_{0}(\alpha)$ for the ground state with $J_{f}=1 / 2$ is mainly the $\alpha=110$ component due to the GT transition of $\mathrm{p}_{1 / 2}(\mathrm{n}) \rightarrow \mathrm{p}_{3 / 2}(\mathrm{p})$, the $\sigma_{2}(\alpha)$ for the 175 keV state with $J_{f}=5 / 2$ is mainly the $\alpha=132$ component due to the T3 transition for the $\mathrm{f}_{5 / 2}(\mathrm{n}) \rightarrow \mathrm{p}_{3 / 2}(\mathrm{p})$ and $\sigma_{3}(\alpha)$ for the 500 keV state with $J_{f}=3 / 2$ includes the $\alpha=110$ component due to the GT and T1 transitions and the $\alpha=122$ component due to the T2 transition.

The ISM GT and ISM T1 NMEs with the quenching coefficient of $k=0.74$ are shown in Table \ref{tab:nme}. The value of $k M_{0}(\mathrm{ISMGT})=0.585$ is the EC NME as already discussed. The $k M_{0}(\mathrm{ISMT1})=0.104$ is much smaller than the value for $k M_{0}$ (ISMGT), being consistent with the major GT component for the ground state NME. On the other hand, $M_{1}$ (ISMT1) is as large as 0.454 and $M_{1}$ (ISMGT) is very small. This supports the main $l$-forbidden $\mathrm{f}_{5 / 2}(\mathrm{n}) \rightarrow \mathrm{p} 3 / 2(\mathrm{p})$ transition for the 1st excited state. Both $M_{2}$ (ISMGT) and $M_{2}$ (ISMT1) are quite small for the 2nd excited state, suggesting the non-QP, the non-$l$-forbidden, but the quadrupole-phonon transition.

On the basis of these findings, the GT NME $M_{i}(\mathrm{GT})$ is evaluated by using the CER NME of $M_{i}(110)$ and the ISM NME of $M_{i}$ (ISMT1) as
\begin{equation}
M_{i}(\mathrm{GT})=M_{i}(110)-\epsilon k M_{i}(\mathrm{ISMT1})
\end{equation}
where the second term is the correction for the tensor contribution with $\epsilon$ being the mixing coefficient. Using the $\beta$-NME of $M_{0}(\mathrm{GT} ; \beta)=0.087$ and the observed CER NME of $M_{0}(110)=0.15 \pm 0.01$ \citep{Ejiri2012} and the ISM NME of $M(\mathrm{ISMT1})=0.24$ for the ${ }^{69}\mathrm{Ga}~3 / 2 \rightarrow 5 / 2$ transition and $k=0.74$, one gets $\epsilon=0.26 \pm 0.06$. Here we assumed the same phase for $\mathrm{M}(\mathrm{GT})$ and $\mathrm{M}(\mathrm{T} 1)$ as given in the ISM. The value for $\epsilon$ is a little smaller than the values used in \citep{Haxton1998, Kostensalo2019}. The obtained GT NMEs for ${ }^{71} \mathrm{Ge}$ are shown in Table \ref{tab:nme}. 

The large absolute value of $M_{0}(\mathrm{GT})=0.603 \pm 0.02$ for the $1 / 2$ ground state agrees within the error with the EC value of 0.585, i.e. the ISM $M_{0}(\mathrm{ISMGT})$ with $g_{\rm A}^{eff}$, reflecting the main spin-stretched GT NME for the $\mathrm{p} 1 / 2$ neutron $\rightarrow \mathrm{p} 3 / 2$ proton transition. The $M_{2}(\mathrm{GT})=0.255 \pm 0.03$ for the second excited state is a half of the GT NME for the $1 / 2$ ground state, due to the admixture of the quadrupole phonon and non-spinstreched transitions. The small $M_{1}(\mathrm{GT})=0.01 \pm 0.05$ for the $5 / 2$ first excited state suggests the main $l$-forbidden transition of $\mathrm{p} 3 / 2 \rightarrow \mathrm{f} 5 / 2$. Actually the NMEs of $l$-forbidden $3 / 2 \leftrightarrow 5 / 2$ transitions in the Ga mass region with $A \approx 70$ are $M(\mathrm{GT})=0.089$ for ${ }^{67} \mathrm{Cu}, M(\mathrm{GT})=0.088$ for ${ }^{69} \mathrm{Ge}$ and $M(\mathrm{GT})=0.093$ for ${ }^{69} \mathrm{Cu}$ \citep{Mo1983, Nesaraja2014, Singh2019}, being consistent with the present $M(\mathrm{GT}) \leq 0.06$ for the ${ }^{71} \mathrm{Ga}$. They are indeed an order of magnitude smaller than the NMEs $M(G T) \approx 0.6$ for the $l$ allowed $3 / 2 \leftrightarrow 1 / 2$ GT transitions.

\section{Nuclear responses for the solar and $^{51} \mathrm{Cr}$ neutrinos}
The solar- $\nu$ rate $R$ (solar- $\nu$ ) is given by using the rate $R_{0}$ (solar- $\nu$ ) for the ground state as \citep{Haxton1998}

\begin{equation}
R(\text { solar }-\nu)=K R_{0}(\text { solar }-\nu), \quad K=1+K_{1}+K_{2} 
\end{equation}

\begin{equation}
K_{1}=0.667 \frac{B_{1}(G T)}{B_{0}(G T)}, \quad K_{2}=0.218 \frac{B_{2}(G T)}{B_{0}(G T)}
\end{equation}

where $K_{i}$ is the relative contribution from the $i^{\text{th}}$ excited states.

\begin{figure}[h]
\hspace{-2cm}
    \centering
    \includegraphics[width=0.35\textwidth]{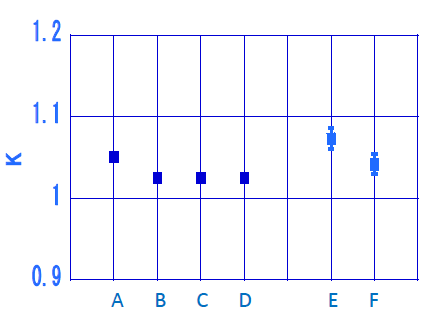}
    \caption{Solar $\nu$ rates for the low lying states in ${ }^{71} \mathrm{Ga}$. The ratios of $K$ to the ground state are plotted. A: estimate \citep{Bahcall1997}, B: ISM in \citep{Haxton1998}. C: ISM in \citep{Kostensalo2019}. D: ISM (present). E: CER (present). F: CER-ISM (present).}
    \label{fig:solar_v_rates}
\end{figure}

\begin{table}[h]
    \centering
    \begin{tabular}{|c|c|c|c|c|}
        \hline
        Method & $K_{1}$ & $K_{2}$ & $K$ & Ref. \\
        \hline
        CER & $0.026 \pm 0.005$ & $0.047 \pm 0.010$ & $1.073 \pm 0.012$ & a \\
        CER/ISM & $<0.005$ & $0.042 \pm 0.010$ & $1.042 \pm 0.012$ & b \\
        ISM & 0.022 & 0.003 & 1.025 & c \\
        ISM & 0.022 & 0.003 & 1.025 & \citep{Kostensalo2019} \\
        pn & 0.019 & 0.032 & 1.051 & \citep{Bahcall1997} \\
        ISM & 0.022 & 0.0034 & 1.025 & \citep{Haxton1998} \\
        \hline
    \end{tabular}
    \caption{Solar $\nu$ rates. a: the present values based on the CER without T1. b: the present values based on the CER with T1 correction, c: the present values based on the ISM.}
    \label{tab:solar_v_rates}
\end{table}

The present ISM NMEs gives $K=1.025$, which agrees well with the previous ISM results \citep{Haxton1998, Kostensalo2019}, but is much smaller than the estimate based on the pn CER \citep{Bahcall1997}. The CER result without the T1 correction is $K(C E R)=1.073 \pm 0.012$, which is close to the value in \citep{Frekers2011}. The present value corrected for the T1 contribution is $K(\mathrm{CER} / \mathrm{ISM})=1.042 \pm 0.012$, which is just between the CER without T1 and ISM values.
The ${ }^{51} \mathrm{Cr}-\nu$ interaction rate is given \citep{Krofcheck1985}

\begin{equation}
R(\mathrm{Cr}-\nu)=K R_{0}(\mathrm{Cr}-\nu), \quad K=1+K_{1}+K_{2} 
\end{equation}
\begin{equation}
K_{1}=0.663 \frac{B_{1}(G T)}{B_{0}(G T)} K_{2}=0.221 \frac{B_{2}(G T)}{B_{0}(G T)}
\end{equation}

This is almost the same as for the solar- $\nu$ because of the similar $\nu$ energies. Then one gets $K\left({ }^{51} \mathrm{Cr}\right)=1.042 \pm 0.012$.

\section{Concluding remarks}
The present work shows for the first time the way to extract GT($\tau\sigma$) $\nu$ nuclear responses $B(GT)$ for the transitions from the $^{71}$Ga ground state with 3/2$^-$ to the $^{71}$Ge ground state, the first excited and the second excited states, respectively with 1/2$^-$, 5/2$^-$, and 3/2$^-$, by using the EC data, the CER data and the ISM calculations.

The axial-vector weak coupling of $g_A^{eff}/g_A$ is derived from the ratio of $M$(ECGT)/$M$(ISMGT) for QP $\beta$ transitions, where the GT NMEs $M$(GT) are well described by the ISM. Here the effective $g_A$ for such non-nucleonic effects that are not explicitly included in the ISM is used for other ISM GT and TL NMEs.

The high energy-resolution CERs with medium energy (0.1-0.2 GeV per nucleon) projectiles are used to excite preferentially the GT states. The cross sections for the s-wave tritons at the forward angles are used to get the NME $M$(110) with $J_p$=1, $J_t$=1, $J_r$=0. It is written as $M$(110)=$M$(GT)+$\epsilon$$M$(T1) with T1=$\tau[Y_2\times\sigma]_1$ being the tensor operator. The Tl response gets significant in case of l-forbidden transition. The T1 mixing coefficient $\epsilon$ is obtained experimentally from the CER NME of $M$(110) and $M$(ECGT) for the l-forbidden transition in $^{69}$Ga. It is $\epsilon = 0.26 \pm {0.06}$ . Then $M$(GT)s for the excited states are derived from the CER $M$(110)s being corrected for the T1 contributions based on the ISM $M$(ISMT1)s.

The 3/2$^-$→1/2$^-$ transition, which is mainly the spin-stretched QP transition, is well described by the CER $M$(110) and the ISM $M$(ISMGT) with the $g_A^{eff}$. The 3/2$^-$→5/2$^-$ transition is the 1-forbidden transition, and thus the GT NME of $M$(ISMGT) is small and the tensor NME of $M$(ISMT1) is large. Then the GT NME is derived from $M$(110) by correcting for the T1 contribution with the mixing coefficient. The 3/2$^-$→3/2$^-$ transition, which is the non spin-stretched transition, is mainly the quadrupole phonon transition, and the $M$(GT) is given mainly by the $M$(110).

The contributions to the neutrino cross sections from the 175 keV and 500 keV excited states in $^{71}$Ga are found to be 4.2 ± 1.2\% for both solar-$\nu$ and $^{51}$Cr by re-analyzing the CER data including the tensor contributions. Then the Ga-anomaly is not due to the nuclear cross-sections.

Medium energy CERs  have been used extensively to get GT NMEs. However, the tensor NME of T(1) shows the same s-wave angular distribution as the GT one, and thus the contribution of the T(1), so far, has not been well corrected for. The present letter shows for the first time the way to correct the CER cross-section for the T(1) contribution.

 \section*{Acknowledgements}
The authors thank Prof. D. Frekers and Prof. J. Suhonen for valuable discussions.

%% The Appendices part is started with the command \appendix;
%% appendix sections are then done as normal sections
% \appendix

% \section{Appendix title 1}
% %% \label{}

% \section{Appendix title 2}
% %% \label{}

% \bibliographystyle{elsarticle-harv}
% \bibliography{references}

\end{document}